\documentclass[%
 reprint,
superscriptaddress,
nofootinbib,
 amsmath,amssymb,
 aps,
pra,
]{revtex4-2}

\usepackage{graphicx}
\usepackage{dcolumn}
\usepackage{bm}

\usepackage{graphicx, color}
\usepackage{subfigure}
\usepackage{braket}
\usepackage{float}
\usepackage{placeins}
\usepackage{algorithm}
\usepackage{algpseudocode}
\usepackage{appendix}
\usepackage{chngcntr}
\usepackage [english]{babel}
\usepackage [autostyle, english = american]{csquotes}
\MakeOuterQuote{"}

\definecolor{chromeyellow}{rgb}{1.0, 0.65, 0.0}

\begin{document}

\title{Experimental demonstration of boson sampling as a hardware accelerator for monte carlo integration}

\author{M. Correa Anguita}
\affiliation{MESA+ Institute for Nanotechnology, University of Twente, P.O.box 217, 7500 AE Enschede, The Netherlands}

\author{T. Roelink}
\affiliation{MESA+ Institute for Nanotechnology, University of Twente, P.O.box 217, 7500 AE Enschede, The Netherlands}

\author{S. Marzban}\altaffiliation{Current affiliation: PiCard Systems B.V., Toernooiveld 100, 6525EC, Nijmegen, The Netherlands}
\affiliation{MESA+ Institute for Nanotechnology, University of Twente, P.O.box 217, 7500 AE Enschede, The Netherlands}

\author{W.J. Briels}
\affiliation{MESA+ Institute for Nanotechnology, University of Twente, P.O.box 217, 7500 AE Enschede, The Netherlands}

\author{C. Filippi}
\affiliation{MESA+ Institute for Nanotechnology, University of Twente, P.O.box 217, 7500 AE Enschede, The Netherlands}

\author{J. J. Renema}
\affiliation{MESA+ Institute for Nanotechnology, University of Twente, P.O.box 217, 7500 AE Enschede, The Netherlands}

\date{\today}

\begin{abstract}
We present an experimental demonstration of boson sampling as a hardware accelerator for Monte Carlo integration. Our approach leverages importance sampling to factorize an integrand into a distribution that can be sampled using quantum hardware and a function that can be evaluated classically, enabling hybrid quantum-classical computation. We argue that for certain classes of integrals, this method offers a quantum advantage by efficiently sampling from probability distributions that are hard to simulate classically. We also identify structural criteria that must be satisfied to preserve computational hardness, notably the sensitivity of the classical post-processing function to high-order quantum correlations. To validate our protocol, we implement a proof-of-principle experiment on a programmable photonic platform to compute the first-order energy correction of a three-boson system in a harmonic trap under an Efimov-inspired three-body perturbation. The experimental results are consistent with theoretical predictions and numerical simulations, with deviations explained by photon distinguishability, discretization, and unitary imperfections. Additionally, we provide an error budget quantifying the impact of these same sources of noise. Our work establishes a concrete use case for near-term photonic quantum devices and highlights a viable path toward practical quantum advantage in scientific computing.
\end{abstract}

\maketitle

\section{Introduction}

Quantum computing has the potential to revolutionize our ability to solve computational problems across a wide variety of application domains, including chemistry~\cite{banchi_molecular_2020, li_hybrid_2024, kirsopp_quantum_2022}, cryptography~\cite{shor_algorithms_1994, bernstein_post-quantum_2009}, and machine learning~\cite{lloyd_quantum_2013, schuld_supervised_2018}. However, many of these applications require large-scale, fault-tolerant quantum computers, a development which is still awaiting substantial improvements in both qubit quantity and quality. In the absence of such devices, a central question in the field of quantum computing is whether we can find applications for the quantum computing devices which are currently available. 

These NISQ (Noisy Intermediate-Scale Quantum) devices are characterized by high noise, an absence of error correction, and  system sizes at around a few hundred qubits or qumodes~\cite{preskill_quantum_2018}. Nevertheless, superconducting and photonic quantum devices have been used for demonstrations of a quantum advantage~\cite{zhu_quantum_2022,madsen_quantum_2022, deng_gaussian_2023, morvan_phase_2024}, i.e. a demonstration of the ability of a quantum computer to outperform a classical device at a well-defined computational task. However, these demonstrations, while a good illustration of the potential power of a quantum computer, do not solve problems of practical interest. Despite several promising candidates~\cite{amos_one-shot_2020, kahanamoku-meyer_classically_2022, aaronson_certified_2023, aaronson_verifiable_2024}, it is still an open problem to find an algorithm that can run on near-term hardware that has real-world utility.  

One of the major blocking issues for finding such an algorithm, particularly in photonic devices, has been that many proposed quantum algorithms have failed to properly take into account the sampling nature of the quantum advantage demonstrations. Quantum advantage demonstrations work by sampling from a probability distribution, whose entries are themselves hard to compute, both classically  and using quantum devices, but where these devices can sample efficiently from the distribution, which is strongly believed not to be the case for classical devices. Directly elevating a quantum advantage demonstration into a useful algorithm therefore requires finding a task, where efficiently quantum hardware can sample from a distribution that is classically hard to sample from, and has practical utility. 

In this work, we experimentally demonstrate such an application---proposed in \cite{renema_monte_2024,andersen_using_2025, andersen_estimating_2025} using photonics (boson sampling) as a hardware accelerator for Monte Carlo integration. The central idea is to use importance sampling to represent a factor in the integral as a multidimensional probability distribution in quantum hardware, which that hardware can then sample from efficiently, with the rest of the integral being computed classically.

We provide evidence that a class of integrals exist for which this procedure offers a quantum advantage: first, we show that certain probability distributions can be efficiently programmed on a boson sampler. Next, we identify several classes of integrals which do not offer a quantum advantage. We find cases where a quantum advantage is negated because the integral is classically easy, and cases where it is negated because it is hard even for quantum hardware. We take this as evidence that a sweet spot exists for integrals which offer a quantum advantage. 

Finally, we show an experimental proof-of-principle demonstration of our algorithm by evaluating the first-order perturbation term using the multi-particle Efimov Hamiltonian on a system of hard-shell bosons in a harmonic trap, using a programmable boson sampler.
We compute how these results would change with improvements in the size and quality of our quantum hardware, and we provide an error budget for our experiment. Taking into account known sources of noise—such as partial distinguishability, discretization, and unitary imperfections—the experimental results are quantitatively consistent with both theoretical predictions and numerical simulations, with observed deviations attributable to these noise sources. This establishes the practical feasibility of our protocol, identifies possible experimental improvements, and highlights the potential utility of our approach.


\section{Boson sampling}

We begin by reviewing boson sampling. Boson sampling is a method for demonstrating a quantum advantage, i.e. the ability of a quantum computer to outperform a classical computer, that is based on quantum optics. In recent years, after the initial proposal by Aaronson and Arkhipov in 2011~\cite{aaronson_computational_2010}, boson sampling has transitioned from a theoretical proposal to experimental implementations pushing the bounds of computational limits, both in classical and quantum systems. 

In a boson sampling experiment, $n$ indistinguishable bosons (most commonly photons) are injected into a linear optical network with $m$ modes, described by a unitary matrix $U$. The input is specified by an occupation pattern $\mu_{\mathrm{in}} = (n_1, n_2, \dots, n_m)$, where $n_j$ denotes the number of bosons entering input mode $j$, with $\sum_j n_j = n$. At the output, one records another occupation pattern $\mu_{\mathrm{out}} = (n'_1, n'_2, \dots, n'_m)$ giving the number of bosons detected per mode. The probability of observing a given output pattern is proportional to the squared modulus of the permanent of a submatrix of $U$:

\begin{equation}
P(\mu_{\mathrm{out}}|\mu_{\mathrm{in}}, U) \propto |\mathrm{Perm}(M)|^2,
\end{equation}

where $M$ is obtained by repeating rows and columns of $U$ according to the occupied input and output modes. The permanent of an $n \times n$ matrix $A = (A_{ij})$ is defined as

\begin{equation}
\mathrm{Perm}(A) = \sum_{\sigma \in S_n} \prod_{i=1}^n A_{i,\sigma(i)},    
\end{equation}

where the sum runs over all permutations $\sigma$ of ${1, \dots, n}$. Computing such permanents is a \#P-hard problem for classical computers~\cite{valiant_complexity_1979}. As a result, boson sampling became one of the promising approaches to experimentally show quantum advantage. 

While early experiments were showing the feasibility of boson sampling with small numbers of modes and photons~\cite{broome_photonic_2013, spring_boson_2013, tillmann_experimental_2013}, recent work has succeeded in scaling these systems up to demonstrate quantum computational advantage, due to technological advances in photonics and quantum interference control.

A significant step towards demonstrating quantum computational advantage using boson sampling was made by Zhong et al in 2019~\cite{zhong_quantum_2020}. Their initial experiment showed Gaussian boson sampling with 20 photons in 60 modes, which was later scaled up in Jiuzhang 2.0 and 3.0, reaching up to 255 photons in 144 modes and further advancing the capabilities of the platform ~\cite{zhong_phase-programmable_2021, deng_gaussian_2023}. These large-scale demonstrations highlight both the rapid progress of boson sampling experiments and the attention they have attracted as milestones of quantum advantage. However, they rely on fixed interferometers tailored to this purpose. A complementary direction, pursued in smaller but reconfigurable devices, is to use programmability to explore mappings between boson sampling and practical computational tasks—a theme we return to later in this work.

\section{Algorithm}

\begin{figure}[t]
    \centering
    \includegraphics[width=0.98\linewidth]{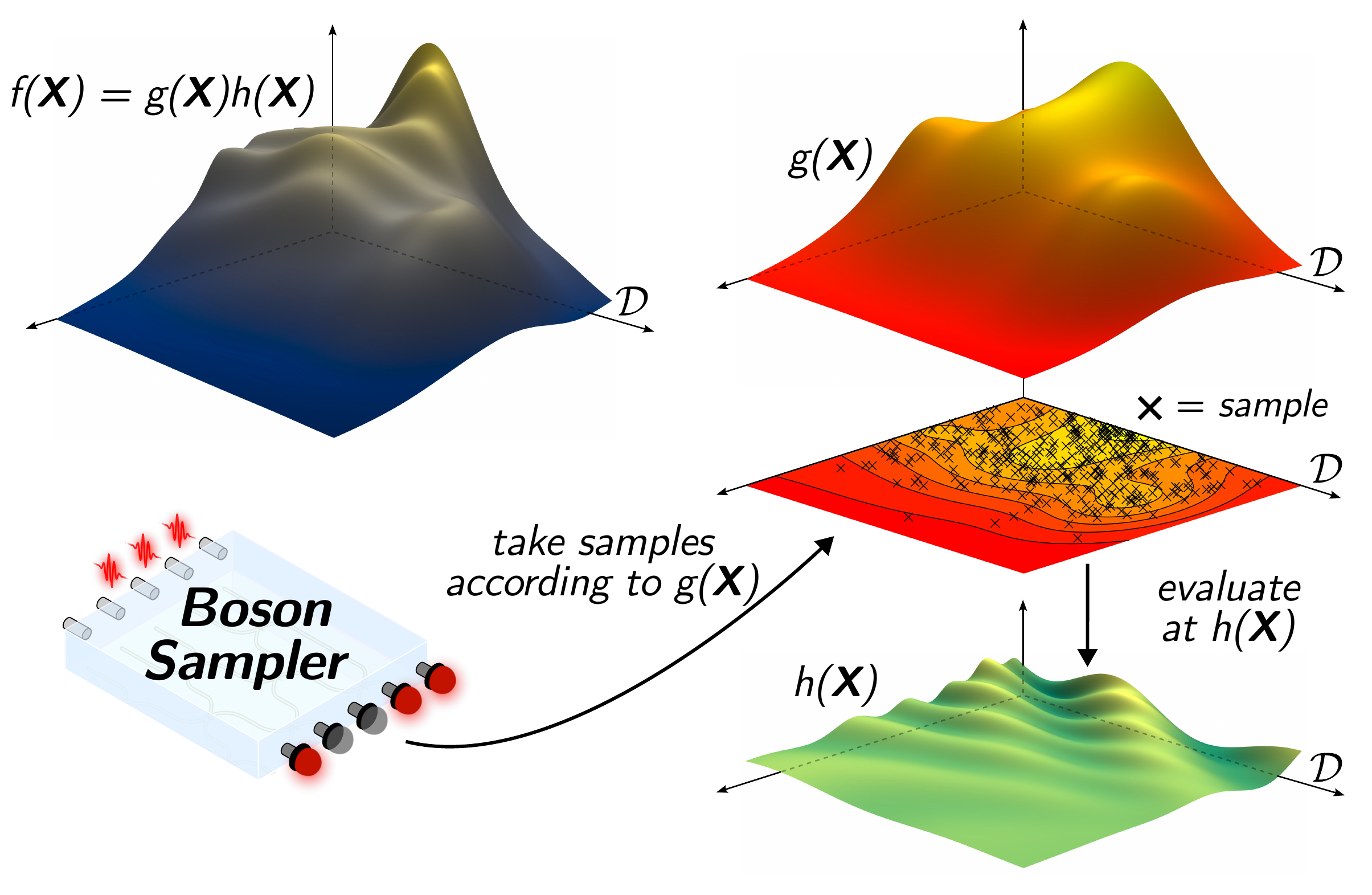}
    \caption{Illustration of Boson-sampling-assisted Monte Carlo integration over a domain $\mathcal{D}$ based on importance sampling. The target function $f(\mathbf{X})$ is factorized into $g(\mathbf{X})$ and $h(\mathbf{X})$. The function $g(\mathbf{X})$ can be implemented in a Boson Sampler, which allows for sampling with a quantum advantage over classical methods. The generated samples are then evaluated using $h(\mathbf{X})$ and summed to efficiently estimate the integral of $f(\mathbf{X})$.}
    \label{fig:BSimportSampling}
\end{figure}

We begin with a high-level overview of the algorithm proposed in \cite{renema_monte_2024, andersen_using_2025, andersen_estimating_2025}. The algorithm centers on importance sampling, a strategy for accelerating Monte Carlo integration. In standard Monte Carlo integration, a quantity of interest is expressed as an integral 

\begin{equation}
F = \int_\mathcal{D} d\mathbf{X}\, f(\mathbf{X}),    
\end{equation}

over a domain $\mathcal{D}$ in some multidimensional space. This can be estimated by drawing samples uniformly from $\mathcal{D}$ and averaging $f(\mathbf{X})$ over those samples. However, a more general and often more efficient approach is to draw a set of samples $\{\mathbf{X}^{(1)}, \mathbf{X}^{(2)},...,\mathbf{X}^{(N)}\}$ from a probability distribution $p(\mathbf{X})$ and rewrite the estimator as
\begin{equation}
    F = \int_\mathcal{D} d\mathbf{X}\, \frac{f(\mathbf{X})}{p(\mathbf{X})} p(\mathbf{X}) \approx \frac{1}{N} \sum_{i=1}^N \frac{f(\mathbf{X}^{(i)})}{p(\mathbf{X}^{(i)})} \quad,
\end{equation}
with $\mathbf{X}^{(i)} \sim p(\mathbf{X})$. 

We take this further by assuming the integrand can be factorized as $f(\mathbf{X}) = g(\mathbf{X}) h(\mathbf{X})$, where $g(\mathbf{X})$ is a probability distribution that can be sampled efficiently using quantum hardware, and $h(\mathbf{X})$ is a classically computable function. This yields a particularly simple estimator:
\begin{equation}
    F = \int_\mathcal{D} d\mathbf{X}\, g(\mathbf{X}) h(\mathbf{X}) \approx \frac{1}{N} \sum_{i=1}^N h(\mathbf{X}^{(i)}) \quad,
\end{equation}
with $\mathbf{X}^{(i)} \sim g(\mathbf{X})$. In our context, $g(\mathbf{X})$ will correspond to the probability distribution generated by a boson sampler, and $h(\mathbf{X})$ will be a function encoding a quantity of interest. The ability to sample from $g(\mathbf{X})$ using photonic hardware enables a hybrid quantum-classical computation in which the sampling is quantum and the evaluation of $h(\mathbf{X})$ is classical. It is useful to note that $\mathbf{X}$ can represent a collection of coordinates, such as the positions of $n$ particles $(\vec{x}_{1}, ...,\vec{x}_n)$ in a $d$-dimensional space. In this case, each particle lives in a single-particle domain $D \subset \mathbb{R}^d$, and the full integration domain is the Cartesian product $\mathcal{D} = D^n$. The integrand $f(\mathbf{X})$ is then a function defined on this $nd$-dimensional space.

In our algorithm, the distribution $g(\mathbf{X})$ is realized by a boson sampler. Each sample $\mathbf{X}^{(i)}$ corresponds to an instance of output occupation pattern $\mu_\text{out}$ of bosons across the interferometer modes, drawn from the boson sampling output distribution. The evaluation of $h(\mathbf{X})$ is performed classically, giving rise to a hybrid quantum–classical computation. Since boson samplers operate over a finite number of optical modes, the single-particle domain $D$ must be discretized into a grid that can be mapped onto those modes. Consequently, the integral over $\mathcal{D} = D^n$ is approximated by a sum over the discretized grid with resolution $\Delta x$. The accuracy of this approximation depends on both the smoothness of $f(\mathbf{X})$ and the fineness of the grid. The discretization introduces a systematic error, in addition to the usual statistical error associated with Monte Carlo sampling. Choosing $\Delta x$ involves a trade-off: smaller values of $\Delta x$ improve resolution but require more optical modes, which may be experimentally costly.

The full boson-sampling-assisted Monte Carlo algorithm is summarized below:

\begin{algorithm}[H]
\begin{algorithmic}
    \caption{Boson-sampling-assisted Monte Carlo integration via importance sampling} \label{alg:bs_mc}
    
    \Require 
    \State - An integral of the form $F = \int f(\mathbf{X})\,d\mathbf{X}$ over a continuous domain $\mathcal{D} = D^n$
    \State - A factorization $f(\mathbf{X}) = g(\mathbf{X}) h(\mathbf{X})$
    \State - A discretization of the domain $D$ with resolution $\Delta x$
    \State - $g(\mathbf{X})$ sampleable using a boson sampler (on the discretized domain)
    \State - $h(\mathbf{X})$ efficiently computable classically

    \Ensure 
    \State - Approximate estimate of $F$ up to error $\mathcal{O}(\Delta x)$

    \State \textbf{Input:} Number of samples $N$
    \State Initialize empty list $H \gets []$
    
    \For{$i = 1$ to $N$}
        \State Draw sample $\mathbf{X}^{(i)} \sim g(\mathbf{X})$ using a boson sampler
        \State Compute $h^{(i)} = h(\mathbf{X}^{(i)})$ classically 
        \State Append $h^{(i)}$ to $H$
        \EndFor

    \State Compute estimate: $F \gets \frac{1}{N} \sum_{i=1}^N h^{(i)}$
    \State \Return $F$
\end{algorithmic}
\end{algorithm}

This strategy is general and applies to a wide range of integrals, provided the factorization is possible and $g(\mathbf{X})$ can be sampled on quantum hardware. In the following section, we demonstrate this approach by applying it to the computation of first-order energy corrections in perturbation theory.

\section{Criteria for quantum hardness}

Finding practical problems which are amenable to  a quantum speedup is a goldilocks problem: the problem must not be too easy to solve classically, since in that case the use of a quantum computer is superfluous, nor should it be so hard that a quantum computer does not help. In our specific setting, the hardness of the problem is governed by the integrand $f(\mathbf{X})$. Without appropriate constraints, the structure of either $g(\mathbf{X})$ or $h(\mathbf{X})$ could enable efficient classical approximations, rendering the quantum sampling protocol ineffective. 

This issue is distinct from the more widely studied forms of reduction in computational hardness caused by experimental imperfections—such as photon loss or partial distinguishability~\cite{aaronson_bosonsampling_2016, shchesnovich2015partial, tichy_stringent_2014, renema_efficient_2018, oszmaniec_classical_2018}—which have been shown to weaken the computational complexity of boson sampling. Here, we focus on a complementary but subtler concern: how the mathematical structure of the functions themselves, even in an idealized setting, can compromise hardness if not properly constrained. In what follows, we introduce two key types of structural restrictions that must be enforced to ensure that neither the sampling distribution nor the classically evaluated function undermines the difficulty of the overall problem.

\subsection{Conditions on Non-Separability}

A useful way to analyze the structure of boson sampling output distributions is through their marginal distributions. A $k$-photon marginal refers to the probability of detecting a specific subset of $k$ photons in selected output modes, irrespective of the positions of the remaining $n - k$ photons. A common example is the single-photon marginal, which captures quantities such as the expected number of photons per mode. Importantly, these low-order marginals are efficiently classically computable, and moreover unchanged when quantum interference is removed—for instance, by rendering the particles distinguishable. In other words, low-order marginals do not reflect the features of the sampling process which carry quantum complexity. What distinguishes the quantum regime is the emergence of nontrivial high-order correlations: joint detection probabilities involving large subsets of photons that encode the full complexity of many-body interference. These correlations are believed to underlie the classical intractability of boson sampling.

In this context, the classically evaluated function $h(\mathbf{X})$ must be sensitive to correlations whose order increases with system size in order to preserve computational hardness. If, instead, $h(\mathbf{X})$ depends only on bounded-order marginals—that is, reduced distributions obtained by tracing out all but a fixed number $k$ of modes—then classical mock-up samplers can reproduce those statistics efficiently \cite{tichy2015sampling, neville_classical_2017}. In such cases, the estimation task no longer probes the genuinely many-body interference that underlies the hardness of boson sampling, and the computational advantage vanishes. The cost of these mock-up samplers scales with the marginal order $k$, requiring evaluation of permanents of size $k$, and thus remains efficient as long as $k$ does not grow with the system size.

\subsection{Conditions on (non-) flatness}

The overall shape of the function $h(\mathbf{X})$ affects classical simulability. We can find two extremal cases: if $h(\mathbf{X}) = \delta(\mathbf{X},\mathbf{X}_0)$ for some $\mathbf{X}_0 \in \mathcal{D}$, then $\int f(\mathbf{X}) d\mathbf{X} = |\mathrm{Perm}(M_{X_0})|^2$, and since computing permanents is believed to be hard for quantum computers, we do not expect a quantum advantage in this case. In the other extreme, if $h(\mathbf{X}) = 1$, we recover the normalization of the boson sampling probability distribution $\int f(\mathbf{X}) d\mathbf{X} = 1$. This builds the intuition that a quantum advantage may arise for those functions which have sufficient features to depend on the integrand in a meaningful way, but not by too much, illustrating the goldilocks problem. 

There are also constraints on the structure of $g(\mathbf{X})$: suppose the sampling distribution $g(\mathbf{X})$ can be approximately partitioned into a collection of disjoint subsets $\{B_k\}_{k=1}^K$ such that within each bin $B_k$, the values of $g(\mathbf{X})$ are nearly uniform:
\begin{equation}
\forall\, \mathbf{X}, \mathbf{X}' \in B_k: \, |g(\mathbf{X}) - g(\mathbf{X}')| \leq \varepsilon
\end{equation}
for some small additive tolerance $\varepsilon$. Then the total probability mass of each bin,
\begin{equation}
G_k := \sum_{\mathbf{X} \in B_k} g(\mathbf{X}),
\end{equation}
can be estimated up to additive error $\delta$ in $\mathcal{O}(n^2/\delta^ 2)$ time using Gurvits algorithm~\cite{gurvits_2002_permanent_approx} applied to the lumped outcomes. Since this quantity aggregates many terms of similar value, the variance of the estimator is suppressed, and additive precision on the order of $\delta = \mathrm{poly}^{-1}(n)$ suffices.

If the number of bins is polynomial, $K = \mathrm{poly}(n)$, and each $G_k$ is estimated to additive error $\delta$, then the resulting coarse-grained approximation $g_{\text{approx}}(\mathbf{X})$—defined by uniformly redistributing $G_k$ over $B_k$—satisfies the total variation bound:
\begin{equation}
\mathrm{TVD}(g, g_{\text{approx}}) \leq \sum_{k=1}^K \left( \varepsilon |B_k| + \delta \right) = \mathrm{poly}(n) \cdot (\varepsilon + \delta).
\end{equation}

This implies that the full distribution $g(\mathbf{X})$ admits a classically efficient approximation in total variation distance whenever it displays approximate piecewise flatness. The insight that lumped or grouped outcome bins open pathways for classical approximability is echoed in recent binning-based strategies for validating and simulating boson sampling experiments~\cite{singh_proof--work_2025, seron_efficient_2024}. These approaches illustrate how coarse-graining the output space can, under certain conditions, reduce the effective complexity of the sampling task.

\begin{figure*}
    \centering
    \includegraphics[width=0.98\linewidth]{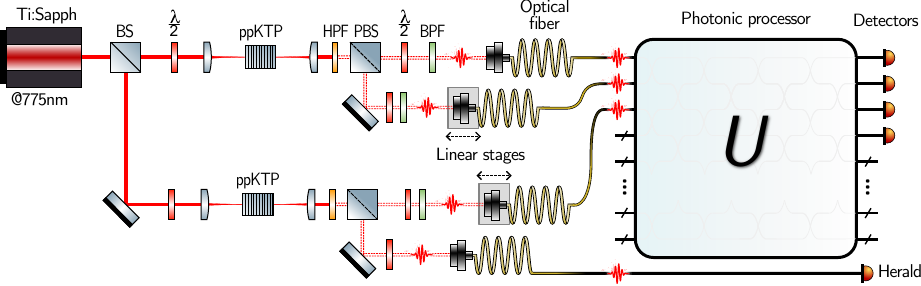}
    \caption{A Ti:Sapphire laser pumps two parallel single-photon sources. The beam is focused onto a periodically poled KTP (ppKTP) crystal, generating photon pairs (red wave packets) via type-II spontaneous parametric down-conversion (SPDC), followed by filtering through a high-pass filter (HPF). The generated photons are separated at a polarizing beam splitter (PBS), further filtered with bandpass filters (BPF), and then coupled into optical fibers. Linear stages are used to control the path lengths, adjusting the relative delays to modulate photon distinguishability. Three photons are directed into the photonic chip for the experiment, while the fourth serves as a herald. A 12-mode unitary transformation $U$ is implemented in the chip. The output photons are measured using superconducting nanowire single-photon detectors (SNSPDs)}
    \label{fig:setup}
\end{figure*}

\section{Experimental implementation}

We present an experimental realization of this algorithm using a boson sampler implemented on an integrated photonic platform. In this section, we first describe the experimental setup, including the photon sources, photonic processor, and detection system. We then detail how a specific quantum problem—computing first-order energy corrections due to an Efimov potential in a harmonic potential—is mapped onto this platform, enabling a demonstration of boson-sampling-assisted Monte Carlo integration in a physically meaningful context.

\subsection{Experimental setup}
 
The experimental setup outlined in Fig.~\ref{fig:setup} is designed to enable high-precision experiments into multiphoton quantum interference, specifically in the context of boson sampling experiments aimed at accelerating Monte Carlo integration. It comprises three main components: photon sources, a tuneable integrated interferometer and single-photon detectors. Our set of single-photon sources is based on periodically poled potassium titanyl phosphate (ppKTP) crystals which generates high-purity single photons at telecom wavelengths, with the setup allowing for tunable degrees of partial distinguishability. These photons are then injected into a large-scale linear optical interferometer, implemented within an integrated photonic processor using silicon nitride waveguides. Finally, the interferometer's outputs are detected by a bank of superconducting nanowire single-photon detectors (SNSPDs).

Our photon sources consist of a pair of periodically poled potassium titanyl phosphate (ppKTP) crystals configured in a Type-II degenerate setup, down-converting light from pulses of a titanium-sapphire (Ti:Sapph) pump laser centered at $775$nm to pairs of single photons at $1550$nm \cite{evans_2010_Phys.Rev.Lett.}, with an output bandwidth of approximately $\Delta \lambda = 20$nm. To enhance the purity of the two-photon state, the photons are filtered using bandpass filters (BPF) with a bandwidth of $\Delta \lambda = 12$nm. We use a single external herald detector and, by conditioning on the detection of three photons after the chip, we post-select on observing the input occupation pattern $\mu_\mathrm{in}= (1,1,1,0, \dots ,0)$ \cite{tillmann_2013}, where the internal degrees of freedom are omitted in the notation for clarity. The photons are then coupled into optical fibers and directed to the photonic processor. By adjusting the relative arrival times of the photons using linear stages on the fiber couplers, we can continuously control their degree of distinguishability. On-chip measurements using the Hong-Ou-Mandel (HOM) effect \cite{hong_measurement_1987} provide a set of calibration measurements to infer the wave function overlap between photons $s_{ij} = \braket{\phi_i | \phi_j}$, where $\ket{\phi_i}$ represents the wave function of photon in input spatial-mode $i$. The maximum wave function overlap between photons, related to the HOM dip visibility via $V = s^2$ was measured.  After filtering, we measure visibilities of 98\%, 95\% and 90\% for photons pairs 1\&2, 1\&3 and 2\&3, respectively (number refers to input spatial mode). It is important to note that HOM tests only provide access to $|s_{ij}|^2$. Therefore, we make the additional assumption that $s_{ij}$ is real. While this assumption may not hold in general, it applies to cases where partial distinguishability is introduced by time-delays, as verified in \cite{menssen2017distinguishability,rodari2024semideviceindependentcharacterizationmultiphoton}.

Our photonic processor consists of an interferometer implemented using silicon nitride waveguides \cite{Triplex, QuiX2021}, with a total of $m = 12$ modes and an optical insertion loss (coupling plus propagation losses) of approximately 5dB (68\%) on average across the input channels. Reconfigurability is achieved through an arrangement of unit cells, each consisting of pairwise mode interactions realized as tunable Mach-Zehnder interferometers \cite{clements_optimal_2016}, adjusted via the thermo-optic effect. For a complete 12-mode transformation, the average amplitude fidelity for randomly chosen unitaries is $\mathcal{F} = {m}^{-1} \mathrm{Tr} (|U^{\dagger}_{\rm set}||U_{\rm get}|) = 90.4 \pm 2.4\%$, where $U_{\rm set}$ and $U_{\rm get}$ are the target and implemented transformations, respectively. However, for highly structured transformations (such as the identity), fidelities as high as 99\% are achievable. The transformation used in this work possesses significant structure—though not to the extent of the identity—and is thus expected to exhibit a fidelity lying between these two regimes. The processor also preserves the second-order coherence of the photons \cite{QuiX2021}.

Photon detection is performed using a bank of superconducting nanowire single-photon detectors (SNSPDs) \cite{ReviewSNSPD, Marsili-SNSPD}, with standard correlation electronics for readout. One click detector is assigned to each output mode, with an additional detector used to herald successful photon generation events. We postselect on triple-coincidence detection events and correct the resulting count rates to account for relative detection efficiency differences across the detector set. An additional noise source arises from higher-order pair generation combined with optical losses, which can lead to spurious postselected events. The pump power of 5mW is chosen to balance photon generation rate and noise suppression, with this effect estimated to contribute less than 1\% to the relative error—rendering it negligible in our analysis.

\subsection{Mapping to harmonic potential system}

\begin{figure}[t]
    \centering
    \includegraphics[width=0.98\linewidth]{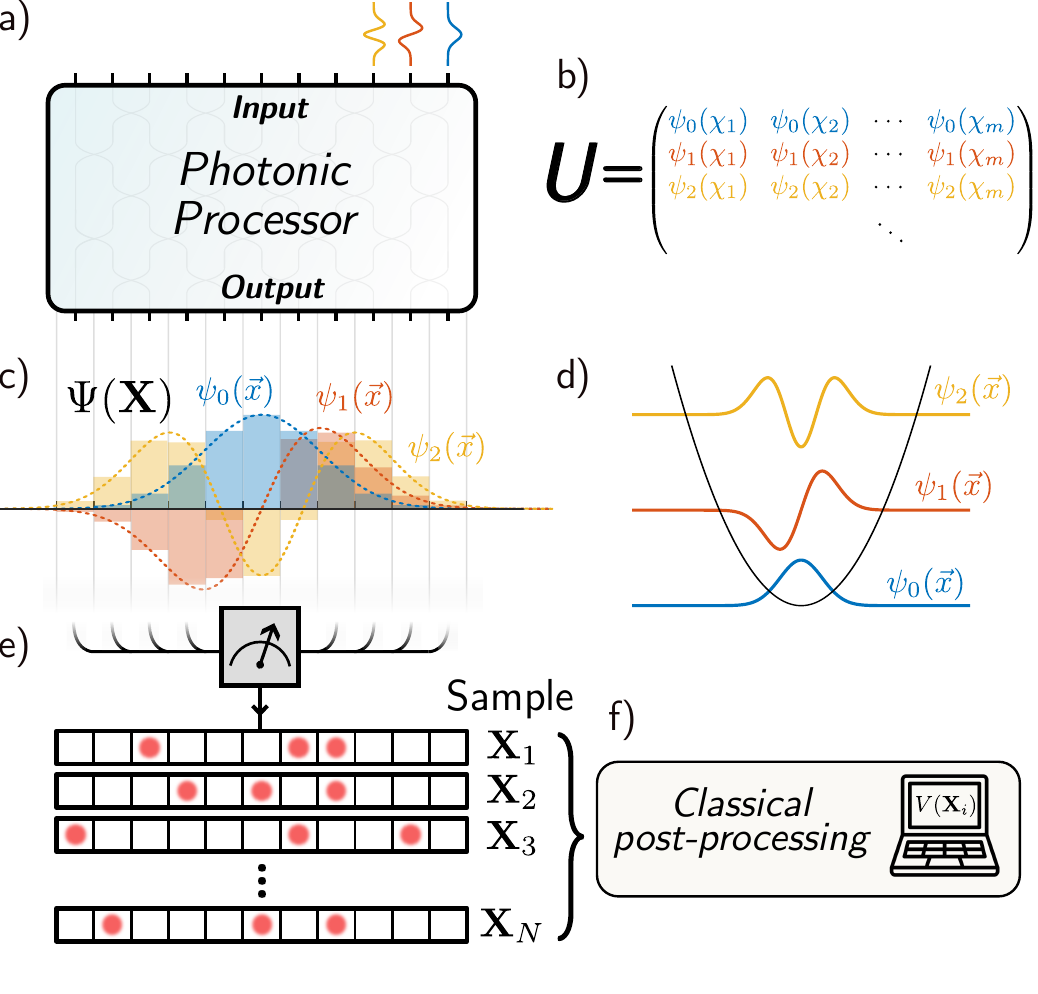}
    \caption{Mapping Boson-Sampling-accelerated Monte Carlo integration to first-order perturbation theory in a harmonic potential. The orbitals (eigenfunctions) of the system are encoded as input modes of the photonic processor \textbf{(a)}, while space is discretized and mapped onto the output modes. The transformation between input and output is described by the unitary matrix $U$ \textbf{(b)}, which defines the operation of the photonic chip. The three-particle wavefunction $\Psi$ \textbf{(c)} is constructed as a bosonic symmetrized product of the individual eigenfunctions of a harmonic potential $V(x)$ \textbf{(d)}. The positions of the three bosons are measured, generating samples \textbf{(e)}, which are then used in classical post-processing to estimate first-order energy corrections \textbf{(f)}.}
    \label{fig:harmPotential}
\end{figure}

We demonstrate the applicability of the general algorithm by mapping it to the estimation of first-order energy corrections in perturbation theory—a standard tool for analyzing weakly interacting bosonic systems. We consider the case of many occupied orbitals, which is known to be difficult due the 
symmetrization of the wave function in this problem~\cite{streltsov_general_2006}. In this setting, the total Hamiltonian is written as $H = H_0 + V$, where $H_0=\sum_{k=1}^n h_0(k)$ with eigenstates given by the symmetrized product of ${\psi_i(\vec{x}_k)}$, and $V$ is a small perturbation. This framework has been applied to study energy-level shifts and splitting in harmonically trapped few-body systems \cite{blume_harmonically_2018, craps_energy-level_2019}. The corresponding many-body ground state wavefunction is denoted by $\Psi_0(\mathbf{X})$, where $\mathbf{X} = (\vec{x}_1, \vec{x}_2, \dots, \vec{x}_n) \in \mathcal{D} = D^n$ represents the coordinates of all $n$ bosons. This state is obtained by bosonic symmetrization over the occupied orbitals ${\psi_i(\vec{x})}$, ensuring full permutation symmetry among the particles.

Provided $V$ does not contain derivatives with respect to the coordinates of $D$, the first-order energy correction to a many-body ground state energy $E_0$ is:
\begin{equation}
    E^{(1)} = \langle \Psi_0 | V | \Psi_0 \rangle = \int_\mathcal{D} |\Psi_0(\mathbf{X})|^2\, V(\mathbf{X})\, d\mathbf{X} \quad,
\end{equation}

which is an integral over $f(\mathbf{X}) = |\Psi_0(\mathbf{X})|^2 V(\mathbf{X})$. This naturally matches our algorithmic framework: the sampling distribution is $g(\mathbf{X}) = |\Psi_0(\mathbf{X})|^2$, while the weight function is $h(\mathbf{X}) = V(\mathbf{X})$. The key point is that $\Psi_0(\mathbf{X})|^2$ itself has the characteristic bosonic structure: when expressed in terms of single-particle orbitals $\psi_i(\vec{x})$, the amplitudes $\Psi_0(\mathbf{X})$ are given by permanents of matrices built from these orbitals. Thus the probability distribution $g(\mathbf{X})$ inherits the same permanental form that underlies boson sampling hardness.

To realize this mapping in a quantum photonic device, we discretize the spatial domain into modes on a linear optical chip. The many-body wavefunction $\Psi_0(\mathbf{X})$ is constructed as a bosonic symmetrization over a set of single-particle orbitals, which are chosen to be eigenfunctions of a harmonic potential. These orbitals are encoded into the chip by defining the unitary transformation $U$ such that $U_{ij} = \psi_i(\chi_j)$, where $\chi_j$ is the $j$-th discretized spatial position\footnote{
Since the matrix $U_{ij} = \psi_i(\chi_j)$ is not unitary due to discretization, we construct a nearby unitary approximation using an SVD-based method. The resulting error in output probabilities is estimated to be on the order of $10^{-6}$ and is negligible for our purposes.},
and $\psi_i(\vec{x})$ is the $i^{\mathrm{th}}$-orbital single-particle wavefunction. This construction ensures that the output amplitudes of the photonic circuit correspond to evaluations of the many-body wavefunction over the spatial grid, such that boson sampling from this unitary yields samples distributed according to $|\Psi(\mathbf{X})|^2$.

\begin{figure*}
    \centering
\includegraphics[width=0.98\linewidth]{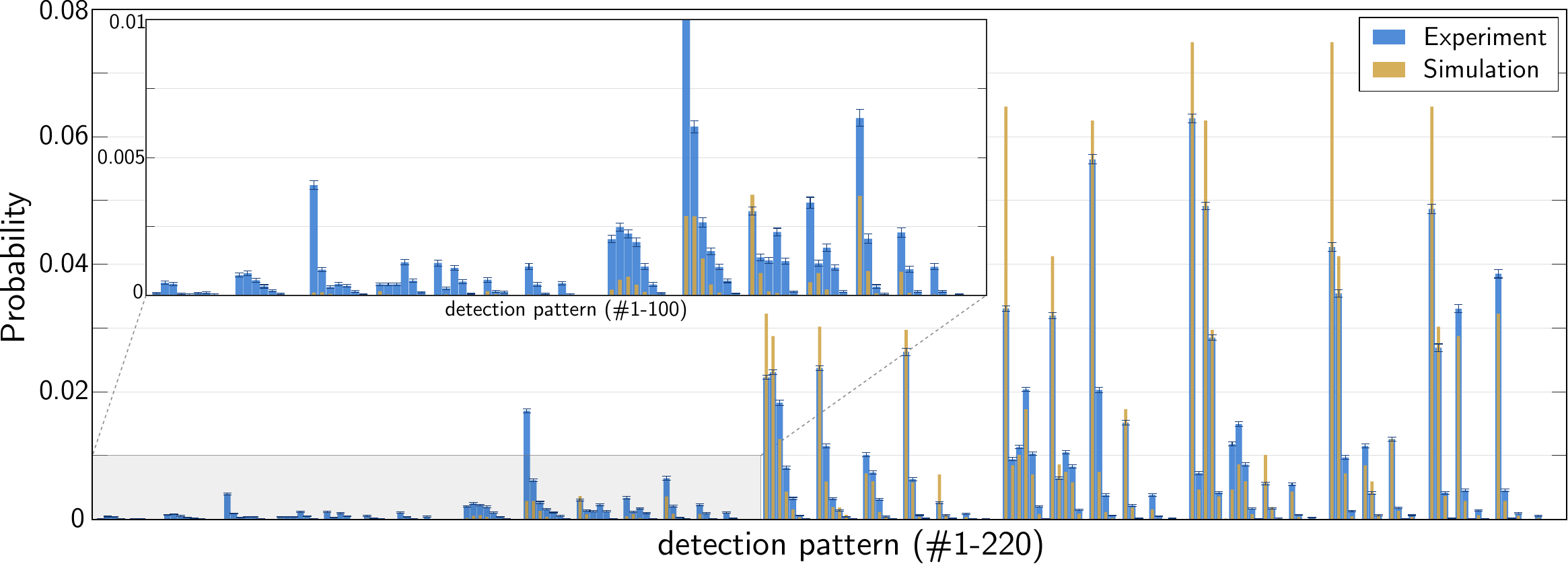}
    \caption{The experimental (blue) and noiseless numerical simulation (gold) probability distributions of photon detection events are shown for every detection pattern (bit string indicating occupied output modes). Data correspond to the near-indistinguishable photon configuration. The x-axis represents distinct multi-photon detection events, while the y-axis indicates their respective probabilities. The inset provides a zoomed-in view of the first 100 detection patterns. Error bars represent statistical uncertainties in the experiment.}
    \label{fig:pdist}
\end{figure*}

Sampling the spatial positions of photons corresponds to drawing samples $\mathbf{X}_i$ from $g(\mathbf{X}) = |\Psi_0(\mathbf{X})|^2$ (Fig.~\ref{fig:harmPotential}e). This process is repeated to build up statistics, and the resulting samples are evaluated classically at the perturbation Hamiltonian $h(\mathbf{X}) = V(\mathbf{X})$ (Fig.~\ref{fig:harmPotential}f), enabling estimation of the first-order energy correction $E^{(1)}$ via Monte Carlo integration.

The perturbation Hamiltonian $V$ is evaluated classically during post-processing and may, in principle, be arbitrary. However, as discussed above, preserving computational hardness requires sensitivity to high-order correlations in the output distribution. If $V$ decomposes into purely single- or two-particle terms, the estimation can be reproduced using partially distinguishable particles and becomes classically tractable. To avoid this, $V$ must involve genuinely many-body structure.

\subsection{Efimov Physics}

A compelling example of a physically relevant perturbation arises in the context of Efimov physics~\cite{naidon_efimov_2017}. In few-body quantum systems, Efimov states emerge when three identical bosons form bound states even in the absence of two-body binding~\cite{efimov_energy_1970, efimov_weakly_1970}. These systems are governed by genuine three-body interactions\footnote{While the characteristic discrete scaling symmetry of the full Efimov spectrum lies beyond our experimental resolution, we leverage the underlying three-body structure as a nontrivial and physically motivated perturbation term.} that cannot be reduced to combinations of single- or two-particle terms—an essential feature for our purposes. 

To introduce a concrete model, we consider an effective three-body potential motivated by Efimov physics:
\begin{equation} 
V_{\mathrm{Ef}}(\mathbf{X}) = -\frac{C + 1/4}{R^2} \quad, 
\end{equation}
where $C$ is a constant and $R^2 = \frac{2}{3}( r_{12}^2 + r_{13}^2 + r_{23}^2)$, with $r_{ij} = |\vec{x}_i - \vec{x}_j|$ denoting pairwise distances. Originating from the universal regime of Efimov physics, this potential captures the essential role of effective three-body interactions. While it is traditionally defined in three dimensions, we apply the same functional form in one dimension to introduce nontrivial three-body correlations. In the context of a proof-of-principle experiment, this provides a well-motivated and physically meaningful perturbation.

An important remark concerns the divergence of the potential as $R\to 0$, which corresponds to the limit where all three particles occupy the same position. In real physical systems, such divergences are typically mitigated by a short-range repulsive interaction that prevents complete spatial overlap. In our experiment, the discretization of space combined with the use of click detectors results in the exclusion of events where two or more photons are detected in the same output mode. This behavior implicitly imposes a hard-shell constraint, with an effective radius determined by the discretization step size. The resulting model—an Efimov-like potential with an additional hard-shell repulsion—defines the actual perturbation used in both our experimental and analytical calculations.

\section{Results}

\subsection{Experimental Results}

The data were collected using a laser with an 80MHz repetition rate and a photon pair generation probability of approximately 0.05\%, yielding a three-photon event rate of roughly ~0.7–1Hz. Each distinguishability configuration was measured in 5-hour batches, repeated across 10 batches. We postselect on events in which exactly three photons are detected, with no more than one photon per output mode—i.e., collision-free three-click events. After correcting for relative detection efficiencies, the total number of recorded three-click events was between $2–3\times10^5$. The results presented here correspond specifically to the near-indistinguishable photon configuration, where quantum interference effects are expected to be strongest.

Figure~\ref{fig:pdist} shows the measured output distribution (blue) compared to a noiseless numerical simulation (gold) for all valid three-photon detection patterns. 
The x-axis indexes the set of allowed detection patterns, while the y-axis shows their corresponding probabilities. Each pattern is represented as a binary string of length 12, where a value of 1 denotes a photon detection in a given mode. The patterns are ordered such that those with detections in lower-numbered modes appear earlier; for example, the pattern (\texttt{111000000000}) appears before (\texttt{110100000000}), and so on. A magnified inset highlights the first 100 patterns to resolve finer structure and low-probability events.

We compare the experimentally estimated output probability distribution with the one obtained by direct numerical computation of the ideal boson sampling circuit. The total variation distance between the two is $\mathrm{TVD}(P_\mathrm{exp}, P_\mathrm{th}) = 0.173$ for indistinguishable photons and $0.110$ for distinguishable photons. These values show that the experimental distribution captures the structured interference pattern expected from the bosonic wavefunction encoded in the photonic chip. Residual discrepancies can be attributed to partial distinguishability and imperfections in the implementation of the target unitary, with the latter appearing to be the dominant source of error, as we will explore in the following subsection. Nevertheless, the observed distribution remains sufficiently accurate to serve as a reliable sampling resource. In the next section, we assess how these samples perform in the context of estimating first-order energy corrections under the Efimov-inspired perturbation potential introduced earlier.

\subsection{Error budget}

In this section, we report the results of the first-order correction $E^{(1)}$ to the harmonic potential energy level of the three-boson state upon introducing the Efimov potential. Table \ref{tab:error_budget} summarizes results from analytical integration (taken as the exact reference), numerical simulations where various sources of noise are introduced progressively, and the experimental measurements. These results help identify the dominant sources of error affecting the experimental implementation of this protocol.

Table \ref{tab:error_budget} shows the computed first-order energy correction $E^{(1)}$ under different levels of noise, using analytical integration, numerical simulations, and experimental data. The "Analytical" entry is obtained via high-accuracy numerical integration with a very fine step size, serving as a baseline reference. "Simulation" entries correspond to numerical simulations of a boson sampling experiment, which remains computationally tractable at these system sizes: we simulate the boson sampler, numerically compute the resulting output probability distribution (pdf), and then plug this simulated pdf into our integration protocol to calculate $E^{(1)}$. "Experiment" entries, by contrast, involve directly measuring the pdf from the actual boson sampling experimental apparatus, and then using this experimentally measured pdf within the same numerical integration protocol.

The parameters listed—$m$, $\bar{s}$, and $F_U$—represent different sources of experimental imperfections. The parameter $m$ indicates the number of optical modes in the boson sampling setup; smaller $m$ values introduce discretization errors, as the continuous spatial domain is represented by fewer discrete modes. The parameter $\bar{s}$ quantifies the average partial distinguishability between photon pairs, serving as a good proxy for the quality of multiphoton interference: lower $\bar{s}$ values indicate increased distinguishability, reducing interference effects and thus degrading the fidelity of the experiment. Finally, the parameter $\mathcal{F}_U$ represents the amplitude fidelity of the implemented photonic unitary transformation, defined as $\mathcal{F}_U = \frac{1}{m} \mathrm{Tr} (| U_\mathrm{ideal}^\dagger|\cdot|U_\mathrm{noisy}|)$, after introducing random Gaussian-distributed noise to each matrix element of the unitary $U$. Lower values of $\mathcal{F}_U$ correspond to larger deviations from the ideal transformation, further decreasing experimental accuracy.

While the effects of partial distinguishability and unitary imperfections are well understood in photonic devices, discretization errors in the MC integrator are more subtle. In particular, care must be taken when applying the hard-shell constraint on a discretized grid. Points that lie exactly on the boundary of the exclusion zone introduce systematic bias: counting them leads to an overestimate of the integral, while discarding them leads to an underestimate. This effect persists even when refining the grid, since the chosen hard-shell distance is 
an integer multiple of
the discretization step. To mitigate this, we introduced a small randomization procedure: for each particle configuration, positions are sampled uniformly within the corresponding mode bins, and the hard-shell condition and Efimov potential are evaluated on these randomized positions. Averaging over multiple random samples per bin yields an unbiased estimate of the integral at a given discretization (see Appendix \ref{app:ControlDiscretizationBias} for details).

\begin{table}[h!]
\centering
\begin{tabular}{|c| c |c|c||c|}
\hline
   \hspace{0.35cm} System \hspace{0.35cm} & \hspace{0.1cm}  m \hspace{0.1cm}   & \hspace{0.25cm} $\bar{s}$  \hspace{0.25cm} & $ \hspace{0.2cm} \mathcal{F}_U$ \hspace{0.2cm} &  \hspace{0.3cm}  $E^{(1)}$  \hspace{0.3cm} \\
    \hline
    \hline
    Analytical & $\infty$   & 1 & $-$   & -0.2453   \\ 
    \hline
    Simulation & 12 & 1     & 100\%     & -0.2467 \\ 
    \hline
    Simulation & 12 & 0.973 & 100\%     & -0.2393 \\ 
    \hline
    Simulation & 12 & 1     & 98.5\%    & -0.2210 \\ 
    \hline
    Simulation & 12 & 0.973 & 98.5\%    & -0.2165 \\ 
    \hline
    Experiment & 12 & 0.973 & $-$       & -0.2114 \\
    \hline
    Simulation & 12 & 0     & 100\%     & -0.1942 \\ 
    \hline
    Experiment & 12 & 0     & $-$       & -0.1806 \\
    \hline
    Simulation & 12 & 0.973 & 90.4\%    & -0.1163 \\ 
    \hline
\end{tabular}
\caption{\textbf{First-order energy correction $E^{(1)}$ under the Efimov potential}, serving as an error budget. From top to bottom, the table presents results with increasing levels of noise. The parameters $m$, $\bar{s}$, and $\mathcal{F}_U$ quantify noise contributions from spatial discretization, boson distinguishability, and photonic chip fidelity, respectively—lower values correspond to higher noise levels. Analytical and ideal simulated results are shown for reference, alongside experimental data and simulations including different noise sources.}
\label{tab:error_budget}
\end{table}

Examining Table \ref{tab:error_budget}, we observe a clear trend: introducing increasing levels of noise systematically shifts the integral toward less negative values. The small residual breaking of monotonicity between the numerical result ($E^{(1)} = -0.2455$) and the noiseless simulation ($E^{(1)} = -0.2467$) can be explained as a consequence of the coarse discretization employed in our experiment. Each mode bin carries the same probability weight, but different random positions within the bin yield slightly different contributions to the integral, an effect amplified by the nonlinearity of the Efimov potential. As expected, this discrepancy decreases as the discretization is refined (see Appendix \ref{app:ControlDiscretizationBias}).

This trend arises because noise reduces quantum bunching: the Efimov integral is dominated by events where multiple particles cluster together, and such clustering is enhanced by multiphoton interference. Imperfections—whether in photon indistinguishability or unitary fidelity—disrupt this interference, leading to more evenly spread configurations and a weaker integral. In this sense, bunching serves as a quantum resource, and any noise erodes it, naturally pushing the result toward less negative values and suppressing false positives.

We now turn our attention to the numerical simulations that incrementally introduce realistic experimental imperfections. By comparing entries where only one source of noise is present, we can assess their individual impact. In particular, we observe that introducing partial distinguishability alone ($\bar{s} = 0.973$, $\mathcal{F}_U = 1$) leads to a moderate shift in the correction ($E^{(1)} = -0.239$), whereas using perfectly indistinguishable photons ($\bar{s} = 1$) with even modest chip infidelity ($\mathcal{F}_U = 0.985$) results in a more pronounced deviation ($E^{(1)} = -0.221$). This indicates that, under these conditions, imperfections in the unitary implementation have a stronger effect on the protocol than reduced multiphoton interference—despite the latter often being regarded as a key challenge in boson sampling.

As noise accumulates further—when both distinguishability and chip infidelity are present—the correction value continues to drift toward zero. The simulation that incorporates both $\bar{s} = 0.973$ and $\mathcal{F}_U = 0.985$ yields $E^{(1)} = -0.217$, and the most extreme simulated case with $\mathcal{F}_U = 0.904$ reaches $E^{(1)} = -0.116$. Our experimental results fall between these two simulated scenarios, which is consistent with expectations. The average chip fidelity of 98.5\% used in the simulations corresponds to identity-like transformations reported by QuiX Quantum, while 90.4\% reflects Haar-random unitaries~\cite{taballione2021universal}. Since our experiment employs an intermediate-depth transformation with some residual structure, it is reasonable that its effective fidelity lies between these two benchmarks, resulting in a measured correction of $E^{(1)} = -0.211$.

Additionally, we can compare the numerical simulation and experimental results in the fully distinguishable regime ($\bar{s} = 0$). The simulated correction for ideal unitary fidelity in this case is $E^{(1)} = -0.194$, while the corresponding experimental result is slightly less negative, at $E^{(1)} = -0.181$. This difference is consistent with the effect of additional chip infidelity in the experimental implementation, further reinforcing the conclusion that limited unitary fidelity is a dominant contributor to the observed discrepancy.

These observations are consistent with the trends seen in the more controlled simulations presented in Appendix~\ref{app:NoiseScaling}. There, we isolate each source of error individually and observe that $E^{(1)}$ degrades more rapidly with unitary infidelity than with partial distinguishability. In particular, moderate reductions in fidelity already produce a noticeable shift in the integral, while $E^{(1)}$ remains relatively stable for small decreases in $\bar{s}$. This reinforces the conclusion that, within the noise regime explored here, imperfections in the implemented unitary dominate over imperfect multiphoton interference. More broadly, these findings underscore the potential and current limitations of using boson sampling devices as hardware accelerators for Monte Carlo integration—offering a promising path forward, provided that key sources of experimental noise can be controlled.

\section{Discussion}

We conclude our work by listing some open problems. 

First, it would be of substantial interest to investigate whether other results on boson sampling besides the ones mentioned here give rise to meaningful constraints the integrands for which a quantum advantage arises. 

Second, it would be of substantial interest to examine which naturally occurring functions have the form required by our method, especially considering alternative implementations of boson sampling that do not use single photons, such as Gaussian boson sampling~\cite{hamilton_gaussian_2017}, which uses squeezed states, superposition sampling~\cite{renema_simulability_2020}, which uses superpositions of single photons and multiple photons, and bipartite boson sampling~\cite{grier_complexity_2022}, which gives rise to permanents of submatrices of non-unitary matrices. 

Third, it would be of substantial interest to investigate connections between the verification problem in boson sampling and the present algorithm. Since our algorithm does not encode its answer into a single or a few outcomes but rather into the set of samples as a whole, it evades the known no-go results on embedding decision problems into a boson sampler~\cite{yung_universal_2019}. 

Fourth, for other many-body quantum mechanical simulations, with applications for example in nuclear physics or quantum chemistry, it would be interesting to see whether sums of permanents (i.e. $P = |\sum_i \mathrm{Perm}(M_i)|^2$ can be represented in a boson sampler. This would allow superpositions of basis states to be probed, allowing variational minimization of the energy cost due to a perturbation term.


\section*{Acknowledgements}
We thank Servaas Kokkelmans for discussions on Efimov physics. This publication is part of the project "At the Quantum Edge" of the research programme VIDI which is financed by the Dutch Research Council (NWO). This project is supported by the Connecting Industries project "Building Einstein’s Dice".  

\section*{Competing interest}
MCA, TR, and JJR are authors of patent WO2024225891A1 on this subject. JJR is a shareholder in QuiX Quantum. 

\appendix
\counterwithin{figure}{section}

\section{Controlling discretization bias in the MC integrator}
\label{app:ControlDiscretizationBias}

While the effects of partial distinguishability and unitary imperfections are well characterized in photonic systems, a subtler challenge arises from the discretization of configuration space in the Monte Carlo (MC) integrator. In particular, care must be taken when enforcing hard-shell constraints on a discretized grid. Due to the sharp cutoff introduced by the hard-shell exclusion zone, points that fall \emph{exactly} on the boundary are ambiguous: should they be considered valid or invalid?

\begin{figure}[b]
    \centering
    \includegraphics[width=0.9\linewidth]{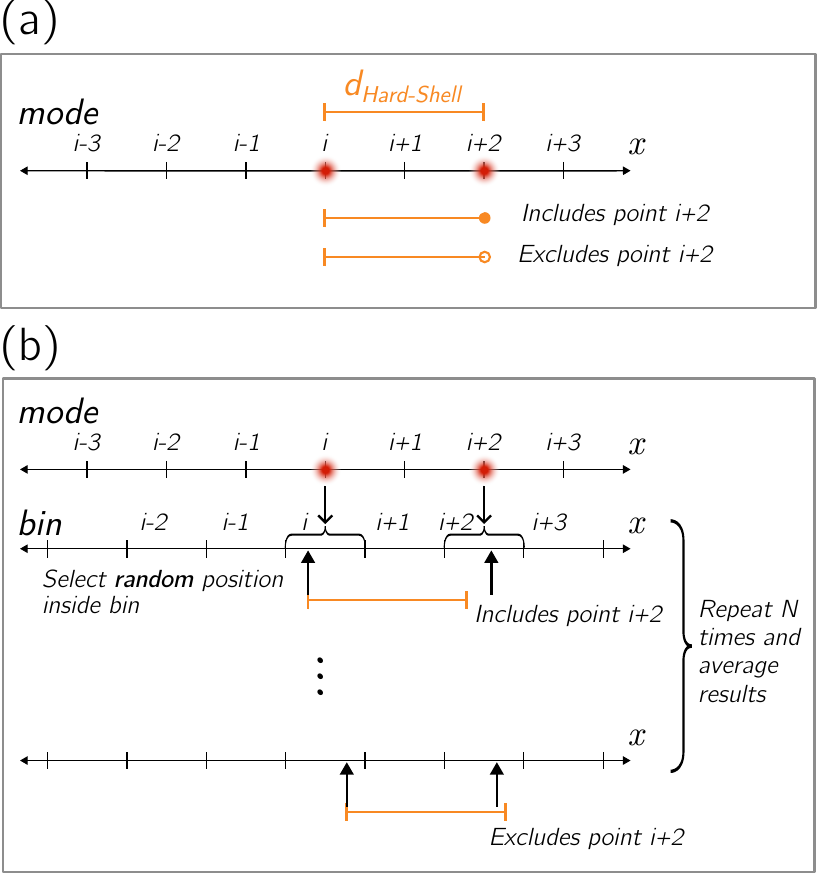}
    \caption{\textbf{(a)} Illustration of boundary ambiguity in deterministic integration. When particle positions are fixed at bin centers, grid alignment effects cause certain points to lie exactly on the boundary of the exclusion zone. Including such points leads to overestimation; excluding them leads to underestimation.
\textbf{(b)} To eliminate this bias, we randomly select positions within the mode bins corresponding to the detected click pattern. The hard-shell constraint is evaluated on these randomized coordinates, and the result is averaged over $N$ repetitions. This procedure removes boundary artifacts and yields an unbiased estimate of the integral.}
    \label{fig:app-randomization concept}
\end{figure}

This ambiguity introduces a systematic bias. Including boundary points leads to an overestimate (in absolute value) of the integral, while excluding them underestimates it. Crucially, this effect does not vanish with finer discretization, because the hard-shell radius remains commensurate with the grid spacing. That is, refining the grid does not remove the alignment artifact — it merely increases the number of points near the problematic boundary. Figure~A.1(a) illustrates this issue schematically.

\begin{figure}[b]
    \centering
    \includegraphics[width=0.98\linewidth]{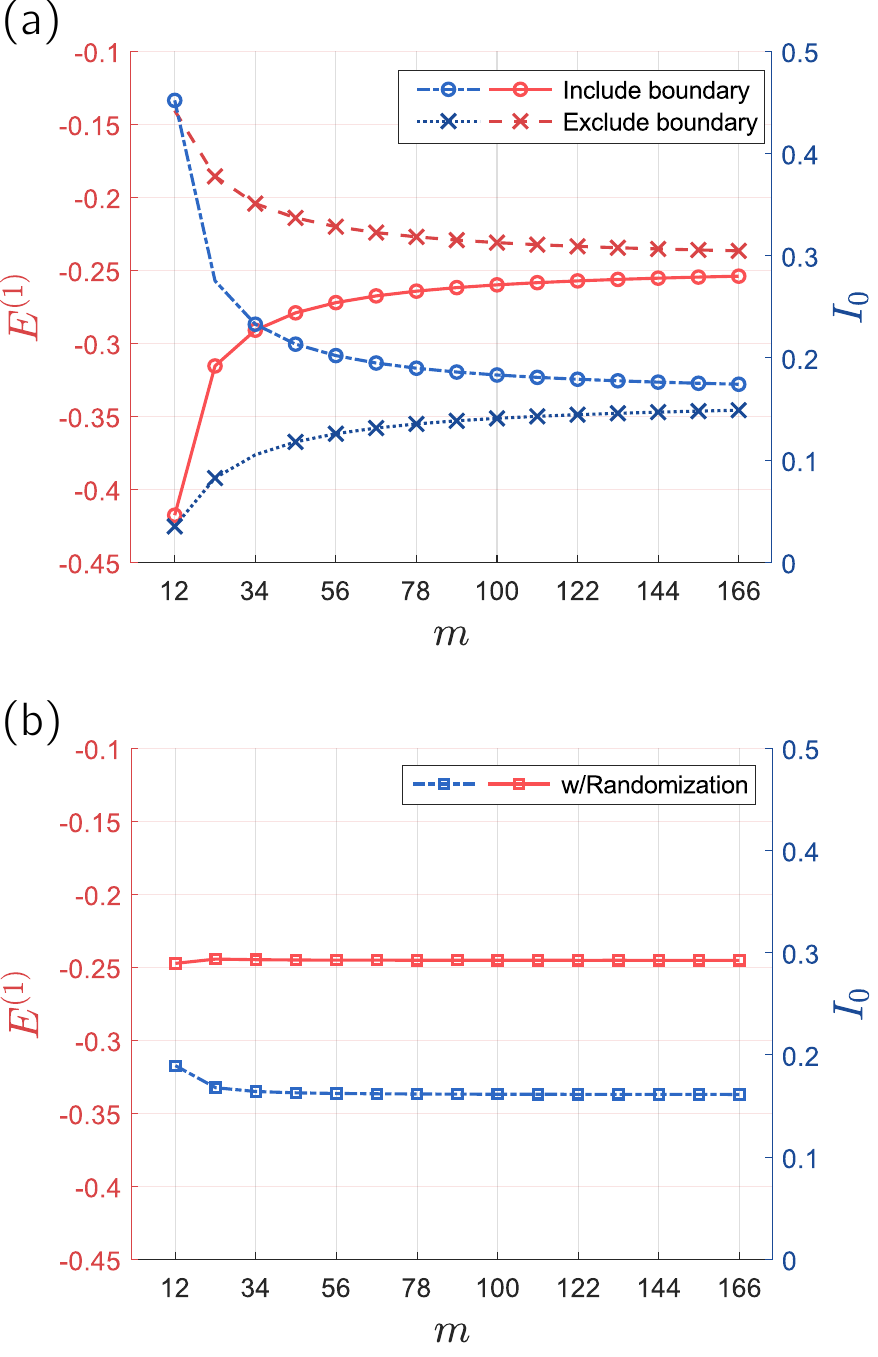}
    \caption{Scaling of the observable $E^{(1)}$ (red, left axis) and the normalization integral $I_0$ (blue, right axis) as functions of the number of modes $m$.
\textbf{(a)} Deterministic integration without randomization. Two conventions are shown: including points on the boundary of the hard-shell exclusion zone (solid and dot-dashed lines), and excluding them (dashed and dotted lines). The resulting bias persists even as $m$ increases.
\textbf{(b)} Integration using the randomization procedure with $N = 10^3$ random positions per bin. Both $E^{(1)}$ and $I_0$ converge cleanly with $m$, and systematic bias is eliminated.}
    \label{fig:app-Int_wWo_randomization}
\end{figure}

To study the impact of discretization bias, we examine how the normalization integral behaves under different treatments of the hard-shell boundary as a function of the number of discretization points (or modes) $m$, using numerical simulations. These simulations are constructed by sequentially adding modes inside each original bin, in a way that keeps the positions of the original $m = 12$ configuration fixed. As a result, the inter-mode spacings are always rational fractions of the original spacing — which we define as the hard-shell exclusion distance. This makes all pairwise distances commensurate with the constraint and amplifies alignment effects, making this construction particularly sensitive to discretization artifacts. In that sense, it serves as a worst-case testbed for evaluating the robustness of the integration method. 

Before looking directly at the effects of these artifacts on $E^{(1)}$, it is convenient to study the behavior of the total probability mass we are integrating over when considering the hard-shell potential. For this purpose, we define $I_0 = \int  p(\mathbf{X}) \Theta_{\text{HS}}(\mathbf{X})d\mathbf{X}$, where $p(\mathbf{X})$ is the three-photon in a harmonic potential probability density used throughout, and $\Theta_{\text{HS}}(\mathbf{X})$ encodes the hard-shell constraint. On a discretized grid, the value of this integral depends sensitively on whether points lying exactly on the boundary are included or excluded. Figure~A.2(a) shows the scaling of both $E^{(1)}$ and $I_0$ with $m$ under these two conventions. Including boundary points leads to systematic overestimation, while excluding them leads to underestimation. These deviations are structural artifacts of the grid and persist even as $m$ increases.

\begin{figure}[b]
    \centering
    \includegraphics[width=0.99\linewidth]{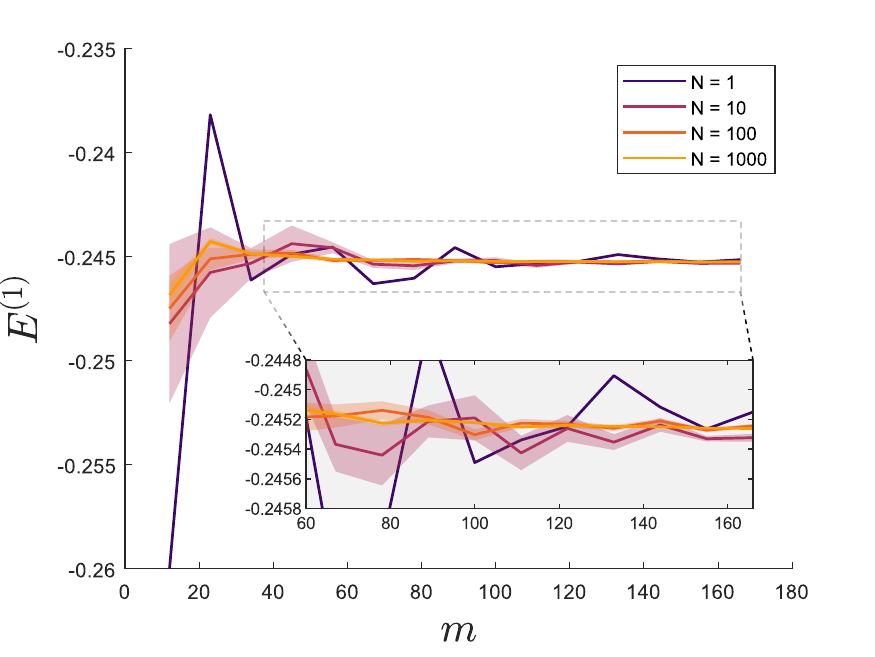}
    \caption{Scaling of the observable $E^{(1)}$ with the number of modes $m$ for different numbers of random samples per bin $N$. As $m$ increases, the curves for different $N$ converge, and the variance decreases. Shaded bands indicate one standard deviation of the mean across repeated runs. The inset zooms into the intermediate-$m$ regime to highlight the convergence behavior.}
    \label{fig:enter-label}
\end{figure}

To mitigate this bias, we implement a simple randomization procedure. For each particle configuration $\mathbf{X}$, positions are drawn uniformly within the corresponding mode bins-the region in space between the midpoint of neighboring points-, and the hard-shell constraint is evaluated on these randomized positions. In the cases where the condition is met, we add the resulting value to $I_0$. The result is averaged over $N$ independent samples per configuration. This “jittering” around the discretization point removes boundary artifacts by probing the entire bin volume, rather than a fixed grid point. FigureA.1(b) shows a schematic of this process, while FigureA.2(b) illustrates its effect: both $E^{(1)}$ and $I_0$ now converge smoothly with $m$ and are have a much flatter response, indicating a fairly significant reduction of bias.

In all simulations, $E^{(1)}$ is normalized by the corresponding value of $I_0$, reflecting the effective probability mass of the allowed region. This ensures consistency across different discretizations, and makes all results comparable on equal footing.\\

Finally, we study how the number of random samples $N$ per bin affects convergence. At small $m$, more samples are needed to suppress variance; at large $m$, even small $N$ suffices. Figure~A3 shows the scaling of $E^{(1)}$ with $m$ for several values of $N$, including shaded bands representing one standard deviation of the mean. As expected, the variance decreases with $N$, and the curves collapse for large $m$, indicating convergence. 

Notably, all curves exhibit a small kink at low $m$, where $E^{(1)}$ becomes slightly more negative before stabilizing. This residual non-monotonic behavior is a consequence of evaluating the Efimov potential on randomized positions within each bin: while each mode bin carries the same probability weight from the underlying pdf, the nonlinear structure of the potential means that different positions within the bin can contribute slightly different values to the integral. At coarse discretization, this effect is amplified; as $m$ increases, the bins shrink and the estimator becomes effectively pointwise, restoring smooth convergence. This behavior explains the small shift between the noiseless and numerical values observed in Table~\ref{tab:error_budget}, and reflects the intrinsic sensitivity of nonlinear integrands to intra-bin variation.\\

All results reported in the main text use this randomized integration strategy.

\begin{figure}[h!]
\begin{subfigure}
    \centering
    \includegraphics[width=0.98\linewidth]{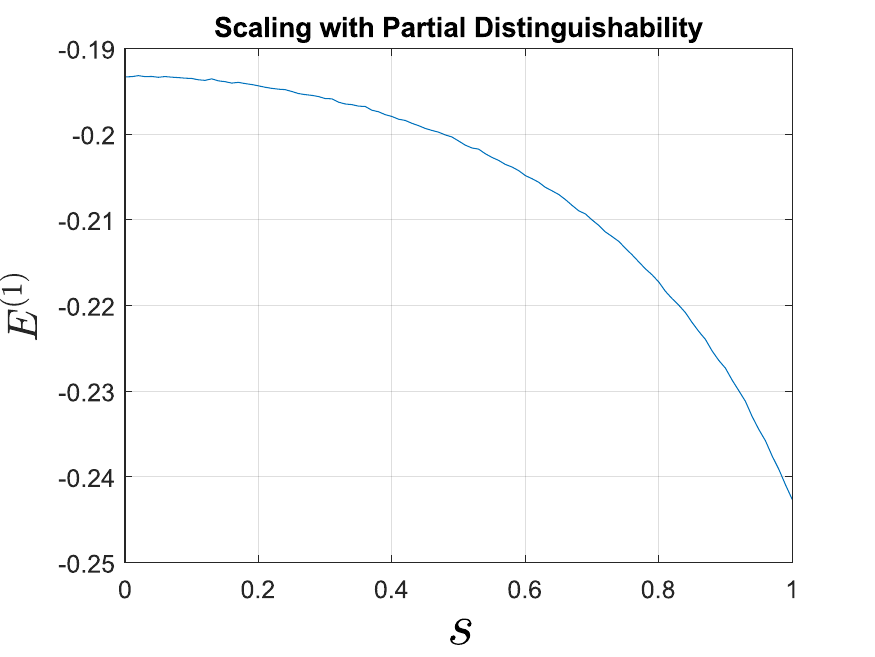}
    \caption{Scaling of the observable $E^{(1)}$ with average photon indistinguishability $s = \langle \phi_i | \phi_j \rangle$, for fixed $m = 12$ and ideal unitary ($\mathcal{F}_U = 1$). A simplified model is used in which all three photon pairs are assigned equal mutual overlap $s$. As indistinguishability increases, quantum interference strengthens, leading to a monotonic decrease in $E^{(1)}$.}
    \label{fig:enter-label}
\end{subfigure}
\vspace{1.7cm}
\begin{subfigure}
    \centering
    \includegraphics[width=0.98\linewidth]{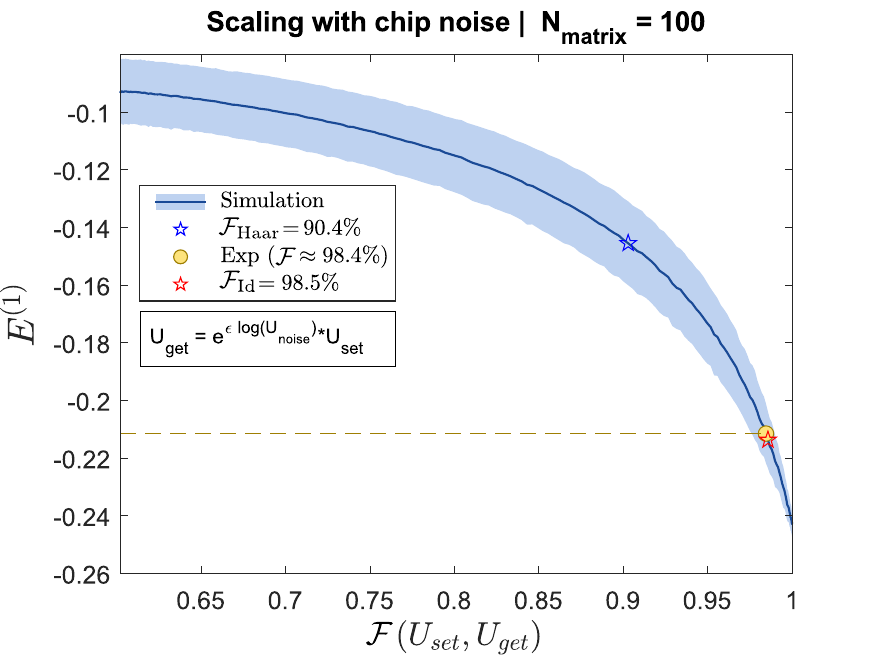}
    \caption{Scaling of the observable $E^{(1)}$ with unitary fidelity $\mathcal{F}(U_{\text{set}}, U_{\text{get}})$, for fixed $m = 12$ and fully indistinguishable photons ($s = 1$). The noisy unitaries $U_{\text{get}}$ are constructed by perturbing a target unitary $U_{\text{set}}$ with random noise matrices, and the noise strength is modulated by a parameter $\epsilon$. The resulting fidelity between $U_{\text{set}}$ and $U_{\text{get}}$ is used as the horizontal axis. Each point represents an average over 100 independently sampled noise realizations, with shaded bands indicating one standard deviation. The yellow circle marks the experimental fidelity observed for the circuit used in this work, while the stars indicate average fidelities reported for identity-like and Haar-random unitaries on the QuiX platform.}
    \label{fig:enter-label}
\end{subfigure}
\end{figure}

\section{Impact of Partial Distinguishability and Unitary Noise}
\label{app:NoiseScaling}

In addition to discretization artifacts discussed in Appendix~\ref{app:ControlDiscretizationBias}, the performance of the Monte Carlo integrator is also affected by noise sources commonly encountered in photonic boson sampling platforms. Here, we focus on two: partial photon distinguishability and unitary imperfections in the implemented circuit. These noise sources degrade the quantum interference that underpins the hardness of the boson sampling problem and affect the observable $E^{(1)}$ accordingly.

To isolate these effects, we perform simulations at fixed discretization resolution ($m = 12$), using the randomized integration procedure described previously. In each case, we vary the strength of the noise while keeping all other parameters fixed.\\

\textbf{Partial Distinguishability:} We model distinguishability by assigning a homogeneous pairwise overlap $s \in [0, 1]$ to all photon pairs. This corresponds to a simplified model where the single-photon states $\phi_i$ have mutual overlaps $\langle \phi_i | \phi_j \rangle = s$ for all $i \ne j$. As distinguishability increases (i.e., $\bar{s} \to 0$), multiphoton interference is suppressed, and $E^{(1)}$ approaches the classical distinguishable limit.

Figure~B1 shows the resulting degradation of $E^{(1)}$ as a function of $\bar{s}$.\\

\textbf{Chip noise:} To model noise in the implemented linear optical circuit, we parameterize the deviation from the ideal unitary ${U}_{\mathrm{set}}$ using the average fidelity $\mathcal{F}_U$. For each value of $\mathcal{F}_U$, we sample a noisy unitary drawn from an ensemble with the corresponding overlap with $U$, and compute $E^{(1)}$ using the randomized integrator. Photon distinguishability is held fixed at $\bar{s} = 1$.

Figure~B2 shows the scaling of $E^{(1)}$ with $\mathcal{F}_U$. The values $\mathcal{F}_U = 98.5\%$ and $\mathcal{F}_U = 90.4\%$ correspond to reported average fidelities for identity-like and Haar-random circuits on the QuiX platform. 

\newpage
\bibliographystyle{unsrt}
\bibliography{refs}

\end{document}